\documentstyle[proceedings,numreferences]{crckapb}
\begin{opening}
\title{STRUCTURE FORMATION:\protect\\
MODELS, DYNAMICS AND STATUS}
\author{T. PADMANABHAN}
\institute{Inter-University Centre for Astronomy and Astrophysics,\\
Post Bag 4, Ganeshkhind, Pune - 411 007, INDIA.\\
{\it Email: paddy@iucaa.ernet.in}}
\end{opening}
\runningtitle{STRUCTURE FORMATION: MODELS, DYNAMICS AND STATUS}
\begin{document}

\def\la{\mathrel{\mathchoice {\vcenter{\offinterlineskip\halign{\hfil
$\displaystyle##$\hfil\cr<\cr\sim\cr}}}
{\vcenter{\offinterlineskip\halign{\hfil$\textstyle##$\hfil\cr<\cr\sim\cr}}}
{\vcenter{\offinterlineskip\halign{\hfil$\scriptstyle##$\hfil\cr<\cr\sim\cr}}}
{\vcenter{\offinterlineskip\halign{\hfil$\scriptscriptstyle##$\hfil\cr<\cr\sim\cr}}}}}

\def\ga{\mathrel{\mathchoice {\vcenter{\offinterlineskip\halign{\hfil
$\displaystyle##$\hfil\cr>\cr\sim\cr}}}
{\vcenter{\offinterlineskip\halign{\hfil$\textstyle##$\hfil\cr>\cr\sim\cr}}}
{\vcenter{\offinterlineskip\halign{\hfil$\scriptstyle##$\hfil\cr>\cr\sim\cr}}}
{\vcenter{\offinterlineskip\halign{\hfil$\scriptscriptstyle##$\hfil\cr>\cr\sim\cr}}}}}

\section {Recipe for the Universe}

All the popular models for structure formation  are based on three key
ingredients: (a) a model for  the background universe  (b)  some
mechanism for generating small perturbations in the early  universe
and (c) specification of the nature of the dark matter.

The background universe is usually taken to be a Friedmann model with
an expansion factor $a(t)$. Such a model is completely specified if
the composition of the energy density and the Hubble constant are
specified. We will take $H_0 = 100 h $~km$s^{-1}$ Mpc$^{-1} $ and
express the energy density of the various species in terms of the
critical density $\rho_c = (3H_0^2/8 \pi G)= 1.88h^2 \times 10^{-29}
gcm^{-3}$, ~by  writing $\rho_i = \Omega_i \rho_c$ for the $i^{th}$
species.  From  various observations, we can impose the  following
constraints: (i) $0.011h^{-2} \la \Omega_B \la  0.016h^{-2}$ ~(ii)
$\Omega_{{\rm vac}} \la 0.8 $~(iii) $\Omega_{{\rm lum}}\simeq
0.007h$~(iv) $\Omega_R = 4.85h^{-2} \times 10^{-5}$~(v) $\Omega _{{\rm
total}} \equiv \Omega \ga 0.3$. Theoretical models strongly favour
$\Omega = 1$ and it is usual to invoke either a cosmological constant
and/or nonbaryonic dark matter to achieve this\cite{IAU117}. We shall
denote by $\Omega_{DM}$ the total contribution due to all nonbaryonic
energy densities.

Models for structure formation also need to assume that small
perturbations in the energy density existed at very early epochs.
These perturbations can then grow via gravitational instability
leading to the structures we see today. In most of the models these
perturbations are generated by processes which are supposed to have
taken place in the {\it very} early universe (say, at $z \ga
10^{18}$). Inflationary models -- which are probably the most
successful ones in this regard -- can produce density perturbations
with an initial power spectrum $P_{{\rm in}}(k) \simeq Ak$. Since each
logarithmic interval in $k$ space will contribute to the energy
density an amount $\triangle_{\rho}^2 (k) \equiv d \sigma^2/d(lnk) =
(k^3 P(k)/2 \pi^2 )$ we find that $ \triangle^2_{\rho} \propto k^4$
for $P \propto k.$ The contribution to gravitational potential from
the same range will be $\triangle^2_{\varphi} = \triangle^2_{\rho}
(9H^4_0/4k^4a^2)$ which is independent of $k$ if $\triangle^2_{\rho}
\propto k^4$. Such a ``scale-invariant''  spectrum is produced in some
other seeded models as well. All these models need to be fine-tuned to
keep the amplitude $A$ of the  fluctuations small  upto, say, $z \ga
10^3$.

Given a Friedmann model with small inhomogeneities  described by a
power spectrum $ P(k, z_{\rm in})$ at a high redshift $z = z_{\rm
in}$, we can predict  {\it unambigiously} the power spectrum  $P(k,
z_D)$ at  the epoch of decoupling $ z_D \approx 10^3$. This is because
the perturbations at all relevant scales are small at $z \ga z_D$ and
we  can use linear perturbation theory during this epoch. The shape of
the spectrum at $z = z_D$  will not be a pure power law since  the
gravitational amplifiction is wavelength-dependent. In general, the
power at small scales is suppressed (relative to that at large scales)
due to various physical processes and  the exact shape of the spectrum
at $z = z_{\rm D} $ depends on the kind of dark matter present in the
universe. In a universe dominated by ``hot dark matter'' particles of
mass $m \simeq 30eV$, the power per logarithmic interval in $k-{\rm
space}, \triangle(k) = (k^3 P(k) / 2 \pi^2)^{1/2}$, is peaked at
$k=k_{\rm max} \equiv 0.11 $ Mpc$^{-1} (m/30 eV) $ and falls
exponentially for $k > k_{\rm max}$. Hence, in these models, the scale
$k = k_{\rm max}$ will go nonlinear first and smaller structures have
to form by fragmentation. If the universe is dominated by ``cold dark
matter'' particles with mass $m \ga 35 $GeV, then $\triangle (k)$ is a
gently increasing function of $k$ for small $k$. If we set $P(k)
\propto k^n$ locally, the  index $n$ changes from  $1 $ ~at~ $ k^{-1}
\ga 200h^{-1} $Mpc ~ to ~$0~ {\rm at}~ k^{-1} \simeq 10h^{-1}$Mpc~ and
to about $(-2 )~{\rm at}~ k^{-1} \simeq 1 h^{-1}$ Mpc. ~In such
models small scales will go nonlinear first and the structure will
develop heirarchically\cite{TPCUP}.

The situation is more complicated if two different kinds of dark
matter are present or if the cosmological constant is nonzero. The
presence of the cosmological constant adds to the power at large
scales but suppresses the growth of perturbations at small scales.
Similar effect takes place if a small fraction of the dark matter is
hot and the bulk of it is cold (eg. $\Omega_{\rm HDM} \simeq 0.2,
{}~\Omega_{CDM} \simeq 0.8)$. In both the cases  there will be more
power at large scales and less power at small scales, compared to
standard CDM model. The spectrum  $\triangle (k)$ is still a gently
increasing function of $k$  and small scales go nonlinear first.

The fact, that one can compute the power spectrum at $z \simeq z_D$
analytically,  allows one to predict large scale anisotropies in CMBR
unambiguously in any given model. Comparing this prediction with the
anisotropy  observed by COBE one can fix the amplitude $A $ of the
power spectrum.  For a wide class\cite{TPDNA92} of the models,
$\triangle (k) \cong 10^{-3} (kL)^2$ with $L \simeq (24 \pm 4 ) h^{-1}
$ Mpc for $k^{-1} \ga 80h^{-1}$ Mpc. For CDM like models the function
$\triangle (k)$ flattens out at larger $k$ and is about unity around
$k^{-1} \simeq 8h^{-1}$ Mpc.  In pure HDM models, $\triangle(k)$ has a
maximum value of $\triangle_m \simeq 0.42h^{-2}(m/30 eV)^2$ at $ k_m
\simeq  0.11 $Mpc$^{-1} (m / 30 eV)$ and decreases exponentially at $k
\ga k_m$.

The evolution of the power spectrum after decoupling (for $z < z_{\rm
D})$ is more difficult to work out theoretically. In general, the
power spectrum grows in amplitude (preserving the shape), as long as
the perturbations are small{\cite{TPCUP}}. In this case, we can write
$\triangle (k, z) = \left[ f(z)/f(z_D) \right] \triangle (k, z_D)$ for
$z < z_D$. For example, in CDM models with $\Omega = 1, f(z) = (1 +
z)^{-1}$; thus $\triangle(k)$ grows by a factor $10^3$ at all scales
between the epoch of decoupling $(z_D \simeq 10^3)$ and  the present
epoch $(z = 0)$, if we assume that linear theory is valid at all
scales. The resulting $\triangle_0 (k),$ obtained  by linear
extrapolation, is often used to specify the properties of the models.
This spectrum correctly describes the power at large scales (say,
$k^{-1} \ga 30h^{-1}$ Mpc) where $\triangle_0 \la 0.1$. The ``density
contrast'' $\sigma(R)$  measures the rms fluctuations in mass within a
randomly placed sphere of radius $R$; upto factors of order unity,
$\sigma(R) \simeq \triangle (k \simeq R^{-1} )$ in heirarchical
models. For most of the, COBE normalised, CDM-like models $\sigma(R)
\approx 1$ around $R \approx 8 h^{-1} Mpc$. Clearly linear theory
cannot be trusted at smaller scales.

There are two major difficulties in understanding the physics at these
small scales. Firstly, the true power  $\triangle_{\rm true}(k)$ of
dark matter  will be larger than $\triangle_0(k)$ due to nonlinear
effects which are difficult to model analytically. Since dark matter
particles interacts only through gravity, it is, of course,  possible
to study the formation of dark matter structures by numerical
simulations.  But to gain insight into the dynamics, it will be
helpful to have  simple analytic models explaining the N-body results.

Secondly, it is important to understand gas dynamical processes before
one can compare theory and observations at small scales. Since baryons
can dissipate energy and sink to the minima of the dark matter
potential wells, the statistical properties of visible galaxies and
dark matter halos could be quite different. The situation is further
complicated by the fact that in hierarchical models, considerable
amount of merging takes place at small scales. It is usual to quantify
our ignorance at these scales by a `bias' (acronym for `Basic
Ignorance of Astrophysical Scenarios') factor $b$ and write $\xi_{\rm
gal} (r) = b^2 \xi_{\rm mass} (r)$. Such a parametrisation is useful
only  if $b$ is idependent of scale and morphology of galaxies.  This
seems to be somewhat unlikely. Since small scale observations are
based on galactic properties, while theoretical calculations usually
deal with underlying mass distribution, any scale (or morphology)
dependence of $b$ could play havoc with predictive power of the
theory.

Recently some amount of progress has been achieved as regards the
first aspect, viz, understanding nonlinear clustering of dark
matter{\cite{HAJS91}. This approach is based on the relationship
between the mean correlation function $\bar \xi (x, a)$ and the mean
relative pair velocity $v (x, a)$. These quantities are related by an
{\it exact} equation.

$${\partial F \over \partial A} - h ( A, X ) {\partial F \over
\partial X} = 0 $$
where $F = \ln \left[ x^3 (1 + \bar \xi) \right], A = \ln a, X = \ln x
$ and $h = - (v / \dot a x )$. The characteristics of this equation
shows that, as the evolution proceeds, power from a large scale $l$ is
transferred to smaller scales upto $x = l (1 + \bar \xi )^{-1/3}$. By
analysing the behaviour of $h$, it is possible to express $\bar \xi
(x, a)$ in terms of the mean correlation function {\it in the linear
theory}, $\bar\xi_L (l, a)$. It turns out that: $\bar\xi (a, x)= Q
[\bar\xi_L (a, l) ] ^n$ with $l^3 = x^3 (1 + \bar\xi)$ where $Q = n =
1$ for $\bar\xi_L \leq 1.2; Q = 0.7, n = 3$ for $1.2 \leq \bar\xi_L
\leq 6.5$ and $Q= 11.7, n = 1.5$ for $\bar\xi_L \geq 6.5$. This
relation shows that ${\bar\xi}$ is steeper than $\bar\xi_L$.

Unfortunately, no such simple pattern exists in the dynamics of
baryons coupled to dark matter. The gas dynamical processes introduce
several characteristic scales into the problem and the evolution
becomes quite complicated. The only reliable way of probing these
systems seems to be through massive hydro simulations which are still
at infancy.

It is clear from the above discussions that our theoretical
understanding is best at large scales $(k^{-1} \ga 30h^{-1 }$ Mpc)
where linear theory is valid, $\triangle_0(k) $ is well determined and
baryonic astrophysical processes are not important. At the
intermediate scales $(3h^{-1}$ Mpc$ \la k^{-1} \la 30h^{-1}$ Mpc), it
is not too difficult to understand the dark matter dynamics by some
approximation but the baryonic physics begins to be nontrivial. At
still smaller scales, $(k^{-1} \la 3h^{-1}$ Mpc) there is considerable
uncertainty in our theoretical predictions. We shall now turn to the
observational probes of the power spectrum at different scales.

\section { Probing the power spectrum}

One of the direct ways of constraining  the models is to estimate the
density contrast $\sigma_{\rm obs}(R)~$ from observations at different
scales and compare it with the theoretically predicted values.
Fortunately, we now have observational probes covering four decades of
scales from $10^{-1}$ Mpc to $10^3$ Mpc. We shall discuss the probes
of different scales in the decreasing order.

\subsection{Near horizon scales:  (300 - 3000) ${\rm h}^{-1}$M{\rm pc}}

These scales are so large that the best way to probe them is by
studying the  MBR anisotropy at angular scales which correspond to
these linear scales. Since a scale $L$ subtends an angle $\theta(L)
\cong  1^{\circ} (L/100h^{-1}$ Mpc) at $z \simeq
z_D$, the $(\triangle T / T)$  observations at $(3^{\circ} -
30^{\circ})$ probe these scales. The COBE-DMR
observations{\cite{SMOOT92}} of $(\triangle T/T)_{\rm rms}$ and
$(\triangle T / T)_Q$ allow one to obtain the following conclusions:
(i) $\sigma (10^3 h^{-1}$  Mpc)$ \simeq 5 \times 10^{-4}$ (ii) The
power spectrum at large scales is consistent with $P_{\rm in}(k)
\simeq Ak$ and, if we take $\Omega = 1, $, then $A^{1/4} \cong (24 \pm
4)h^{-1}$ Mpc  (iii) In this range, $\sigma(R) \cong (24 \pm 4h^{-1}$
Mpc$/R)^2$.

\subsection{ Very large scales : $(80 - 300) h^{-1}$ Mpc}

\subsubsection {CMBR probes:}

These scales span $(0.8^{\circ} - 3^{\circ})$ in the sky at $z \simeq
z_D$. Several ground based and balloon-borne experiments to detect
anisotropy in MBR probe this scale. For example, the UCSB South Pole
experiment has  reported{\cite{SCHUSTJ93}} a preliminary `detection'
of $(\triangle T/T)\simeq 10^{-5}$ at $1.5^{\circ} $ scale, and a 95\%
confidence level bound of $(\triangle T/T) <  5 \times 10^{-5}$. This
translates into the constraint  of $\sigma (10^2h^{-1}$ Mpc) $\la 5
\times 10^{-2}$.

The angular anisotropy of CMBR is dominated by the gravitational
potential wells of dark matter at large scales. However, at $\theta
\simeq 1^{\circ}$ baryonic process affect the pattern of anisotropy
significantly. The precise determination of degree scale anisotropy
can, therefore, help  in distinguishing between different
models{\cite{WHITESCOTT94}}.

\subsubsection{ Galaxy  surveys:} Some galaxy surveys, notably CfA2
survey and pencil-beam surveys probe scales which are about
$10^2h^{-1}$ Mpc in depth{\cite{BROAD90}}. Unfortunately, the
statistics at these large scales is not good enough for one to obtain
$\sigma(R) $ directly from these surveys.

\subsection{ Large scales : ~(40 - 80)~ $h^{-1}$~ Mpc}

\subsubsection{ CMBR probes:}
The scales correspond to $\theta_{MBR} \simeq (24' - 48')$ and are
probed by the experiments looking for small angle anisotropies in MBR.
The claimed detection{\cite{CHENG93}} by MIT-MASM of $(\triangle T/T)
\cong  (0.5 - 1.9) \times 10^{-5}$ at  $\theta \simeq  28', $ if
confirmed, will give a bound of $\sigma (50h^{-1}$ Mpc $) \la  0.3.$

\subsubsection{ Galaxy Surveys:}
Several galaxy surveys, in particular the IRAS-QDOT and APM surveys,
give valuable information about this range{\cite{RR90}}. The angular
correlation of  galaxies, measured by APM survey is $\omega (\theta)
\simeq (1-5) \times 10 ^{-3} $ at $\theta \simeq 14^{\circ} $. This
corresponds to  $\sigma (50 h^{-1}$ Mpc$)\cong 0.2$. What is more
important, these surveys  can provide  valuable information about the
shape of the power spectrum in this range if we assume that galaxies
faithfully trace the underlying mass distribution.

\subsubsection{Large scale velocity field:}

Using distance indicators which are independent of Hubble constant, it
is possible to determine the peculiar velocity field $v(R)$ of
galaxies upto about $80h^{-1}$ Mpc or so. The motion of these galaxies
can be used to map the underlying gravitational potential at these
scales. Careful analysis of observational data shows{\cite{DEKEL94}}
that $v(40h^{-1}$ Mpc$)\simeq (388 \pm 67)$ kms$^{-1}$ and $v
(60h^{-1}$ Mpc$) \simeq (327 \pm 82)$ kms$ ^{-1}$. From these values
it is possible to deduce that $\sigma (50h^{-1}$ Mpc $) \simeq 0.2$.
These observations also allow us to determine the value of the
parameter $(\Omega^{0.6}/b_{\rm IRAS})$ where $b_{\rm IRAS}$ is the
bias factor with respect to IRAS galaxies. One finds that
$(\Omega^{0.6} / b_{\rm IRAS}) = 1.28^{+ 0.75}_{-0.59}$ which implies
that if $\Omega = 1$, then $b_{\rm IRAS} =  0.78 ^{+0.66}_{-0.29}$ and
if $b_{\rm IRAS} = 1$ then $\Omega =  1.51 ^{+1.74}_{-0.97}.$

\subsubsection{Clusters and voids:} The cluster-cluster  corelation
function and the spectrum of voids in the universe can, in principle,
tell us something about these scales. Unfortunately, the observational
uncertainties are so large that one cannot yet  make quantitative
predictions.

\subsection{Intermediate scales : $(8 - 40) h^{-1}$ Mpc}

\subsubsection{ Galaxy Surveys:} The galaxy - galaxy correlation
function $\xi_{\rm gg} \cong [r / 5.4 h^{-1}Mpc]^{-1.8}$ is fairly
well determined at these scales. Direct observations suggest that
$\sigma_{\rm gal} (8h^{-1}$ Mpc$ )\simeq 1$ but  the $\sigma_{\rm DM}$
and $\sigma_{\rm gal}$ at these scales can be quite different because
of possible biasing.

\subsubsection{ Cluster Surveys:} There have been several attempts to
determine the correlation function of clusters of different classes.
It is generally believed that $\xi_{\rm cc} \simeq (r/L)^{-1.8}$ with
$L\simeq 25 h^{-1}$ Mpc. The index $n = 1.8$ is fairly well determined
though the scale $L$ is not; in fact, $L$ seems to depend on the
richness class of the cluster. The quantity $(\xi_{\rm cc}/\xi_{\rm
gg})^{1/2}$ can be thought of as measure of the relative bias between
cluster and galaxy scales. Observations suggest{\cite{DALTON92}} that
this quantity depends on the cluster class and varies in the range $(2
- 8)$. The observational uncertainties are still quite large  for this
quantity to be of real use; but if the observations  improve we will
have valuable information from $\xi_{cc}$.

\subsubsection{ Abundance of rich clusters:} The scale $R=8h^{-1}$ Mpc
contain a mass of $1.2 \times 10^{13} \Omega h^{-1}_{50} M_{\odot}$.
When this scale becomes nonlinear, it will reach an overdensity of
about $\delta \simeq 178$, or -- equivalently -- it will contract to a
radius of $R_f \simeq (8h^{-1}$ Mpc)  $/(178)^{1/3} \simeq 1.5 h^{-1}$
Mpc. A mass of $10^{15}M_{\odot}$ in a radius of $1.5 $ Mpc is a good
representation of Abell clusters we see in the universe. {\it This
implies that the observed abundance of Abell clusters can be directly
related to $\sigma (8 h^{-1}$} Mpc). Several people have attempted to
do this{\cite{SDMWHITE93}}; the final results vary depending on the
modelling of Abell clusters, and give $\sigma (8h^{-1}$ Mpc$) \simeq
(0.5 - 0.7)$. Since $\sigma_{\rm gal}(8h^{-1}$ Mpc)$\simeq 1$, this
shows that $b \simeq (1.23 - 2) $ at $ 8h^{-1}$ Mpc.

It is possible to give this argument in a more general
context{\cite{KSTP94}}. Suppose that the contribution to critical
density from collapsed structures with mass larger than $M$ is
$\Omega(M)$, at a given redshift $z$. Then one can show that

$$ \Omega (M) = erfc \left[ { \delta_c (1 + z) \over \sqrt 2
\sigma_0(M) } \right]$$
where $\delta_c = 1.68$ and erfc(x) is the complementary error
function. The Abell clusters (at $z = 0$) contribute  in the range
$\Omega \simeq  (0.001 - 0.02) $. Even with such a wide uncertainty,
we get  $\sigma_{\rm clus} \simeq (0.5 - 0.7) $.

\subsection{Small scales : $(0.05 - 8) h^{-1}$ Mpc}

These scales correspond to structures with $M_{\rm smooth} \simeq (3
\times 10^8 - 1.2 \times 10^{15}) ~\Omega h^{-1}_{50} M_{\odot}$ and
we have considerable amount of observational data covering these
scales. Unfortunately, it is not easy to make theoretical predictions
at these scales because of nonlinear, gas dynamical, effects.

\subsubsection{ Epoch of galaxy formation:} Observations indicate that
galaxy-like structures have existed even at $z \simeq 3$. This
suggests that there must have been sufficient power at small scales to
initiate galaxy formation at these high redshifts. Unfortunately, we
do not have reliable estimate for the abundance of these objects at
these redshifts and hence we cannot directly use it to constrain
$\sigma (R)$.

\subsubsection{ Abundance of quasars:} The luminosity function of
quasars is fairly well determined upto $z \approx 4$. If the
astrophysical processes leading to quasar formation are known, then
the luminosity function can be used to estimate the abundance of host
objects at these redshifts. Though these processes are somewhat
uncertain, most of the models for quasar  formation suggest  that we
must have $\sigma (0.5h^{-1}$ Mpc$) \ga 3.$

\subsubsection{Absorption systems:} The universe at $1 \la z \la 5$ is
also probed by the absorption of quasar light by intervening objects.
These observations suggest that there exist significant amounts of
clumped material in the universe at these redshifts with neutral
hydrogen column densities of $N_{\rm HI} \simeq     (10^{15} -
10^{22}) $cm$^{-2}.$ We can convert these numbers into abundances of
dark matter halos by making some assumptions about this structure. We
find that{\cite{KSTP94}} in the redshift range of $z \simeq (1.7 -
3.5)$ damped Lyman alpha systems contribute a fractional density of
$\Omega_{Ly} \simeq (0.06 - 0.23).$ This would require
$\sigma(10^{12}M_{\odot}) \simeq (3 - 4.5)$.

\subsubsection{ Gunn-Peterson  bound:}  While we do see absorption due
to {\it clumped } neutral hydrogen, quasar spectra do not show any
absorption due to smoothly distributed neutral hydrogen. Since the
universe  became neutral at $z \la z_D \simeq 10^3$, and since galaxy
formation could not have made all the neutral hydrogen into clumps, we
expect the IGM to have  been ionised sometime during $5 \la z \la
10^3$.  It is not clear what is the source for  these ionising
photons. Several possible scenarios (quasars, massive primordial
stars, decaying particles etc.) have been suggested in the literature
though none of these appears to be completely satisfactory. In all
these scenarios, it is necessary to form structures at $z \ga 5$ so
that an ionising flux of about $J = 10^{-21} $ergs cm$^{-2}$s$^{-1}$
Hz$^{-1}$ sr$^{-1}$ can be generated at these epochs. Once again, it
is difficult to convert this constraint into a firm  bound on $\sigma$
though it seems that $\sigma (0.5 h^{-1}$ Mpc$) \ga 3 $  will be
necessary.

\section {Gravitational lensing and large scale structure}
In the above discussion we have not taken into consideration the
constraints imposed by gravitational lensing effects on the structure
formation models. This aspect will be discussed in detail in the other
articles in this volume; here we shall contend ourselves with a brief
mention of the possibilities.

Gravitational lensing  probes the gravitational potential directly and
can provide valuable information at very different scales. At the
largest scales $(R \simeq 10^3 Mpc)$ lensing can be used to probe the
geometry of the universe. For example, it is possible to put firm
bounds on the energy contributed by cosmological constant from such
considerations.

At intermediate scales ( $R \simeq 50 Mpc)$ lensing has the potential
of providing information about the power spectrum of fluctuations
which are in the quasilinear phase. In principle the distortion of
images can be inverted to obtain this information, though in practice
this is extremely difficult.

At smaller scales,  the ``weak lensing'' -- leading to arcs and
arclets at cluster scales -- is already providing a clue to the
mapping of dark matter distribution in clusters. On the other hand,
direct optical and X-ray  observations provide us information about
the distribution of visible matter in clusters. The combination of
these techniques should give us  valuable information as regards the
dynamical processes which separated baryons from dark matter.

At still smaller scales, galactic potentials have the capacity to
produce multiple images of distant sources. The statistics of these
multiple images depends crucially on the core radii of the galaxies,
which in turn depends sensitively on the structure formation models.
The absence of significant number of multiple images with large
angular separations puts severe constraints on models for structure
formation. The analytic modelling of nonlinear dark matter clustering
described earlier  could be used to strengthen these constraints still
further.

\section{ Scorecard for the models}

The simplest models one can construct will contain a single component
of dark matter, either cold or hot. Such models are ruled out by the
observations. The HDM models, normalised to COBE result will have
maximum power of $\triangle_m \cong 0.42h^{-2} (m/30eV)^2 $ at $k =
k_m =  0.11$ Mpc$^{-1}(m/30eV)$. In such a case, structures could have
started forming only around $(1 + z_c) \cong (\triangle_m / 1.68)
\cong h^{-2}_{50} (m/30eV)^2$ or at $z_c \cong 0 .$ We cannot explain
a host of high-$z$ phenomena with these models. The pure CDM models
face a different difficulty. These models, normalised to COBE, predict
$\sigma_8 \simeq 1$, which is too high compared to the bounds from
cluster abundance. When nonlinear effects are taken into account, one
obtains $\xi_{\rm gg} \propto r^{-2.2}$ for $h = 0.5$ which is too
steep compared to the observed value of $\xi_{\rm gg} \propto
r^{-1.8}$. In other words, CDM models have   wrong shape  for $\xi(r)$
to account for the observations.

The comparison of CDM spectrum with  observations suggests that we
need more power at large scales and less power at small scales. This
is precisely what happens in models with both hot and cold dark matter
or in models with nonzero cosmological constant.  These models have
been extensively studied during the last few years, and they  fare
well as far as large and intermediate scale observations are
concerned. However, they have considerably less power at small scales
compared to CDM model. As a result, they do face
difficulties{\cite{KSTP94}} in explaining the existence of high
redshift objects like quasars. For example, a model with $30\%$ HDM
and $70\%$ CDM will have $\sigma_{0.5} \simeq 1.5$; to explain the
abudnance of quasars comfortably one needs  $\sigma_{0.5} \simeq 3.0$.
To explain the abundance of damped Lyman alpha systems one requires
still larger vlaues of about $\sigma_{0.5} \approx 4 $ or so.
Demanding that $\sigma (10^{12}M_{\odot}) >   2$ [which is equivalent
to saying that $10^{12}M_{\odot}$ objects must have collapsed at a
redshift of $z_{12} = (2 / 1.68) -1 \simeq 0.2]$ will completely rule
out this model. Similar difficulties exist in models with cosmological
constant. Notice that all models are normalised using COBE results at
very large scales. Hence the severest constraints are provided by
observations at smallest scales, since the ``lever-arm'' is longest in
that case.

The comparison of models show that it is not easy to accommodate all
the observtions even by invoking two components to the energy density.
(These models also suffer from serious problems of fine-tuning). By
and large, the half-life of such quick-fix models seem to be about 2-3
years. One is forced to conclude that to make significant progress it
is probably necessary to perform a careful, unprejudiced analysis of:
(a) large scale observational results and possible sources of error
and (b) small scale baryonic astrophysical processes.

\end{document}